\documentstyle[12pt]{article}
\begin{document}
\huge
\begin{center}
\Large{NEW DUAL RELATIONS BETWEEN QUANTUM FIELD THEORY AND STRING REGIMES IN CURVED BACKGROUNDS}\\
\vspace{3cm}
M. RAMON MEDRANO \footnotemark[1]
\footnotetext[1]{Departamento de F\'{\i}sica Te\'orica, Facultad de Ciencias
F\'{\i}sicas, Universidad Complutense, E-28040, Madrid, Spain.} 
and N. SANCHEZ \footnotemark[2]

\vskip 3cm
\begin{abstract}
A ${\cal R}$ "dual" transform is introduced which relates Quantum Field Theory
and String regimes, both in a curved background with D-non compact dimensions. 
This operation maps the characteristic length of one regime into
the other (and, as a consequence, mass domains as well).
The ${\cal R}$-transform is not an assumed or {\it a priori} imposed symmetry
but is revealed by the QFT and String dynamics in curved backgrounds.
The Hawking-Gibbons temperature and the string
maximal or critical temperature are ${\cal R}$-mapped one into the other.
If back reaction of quantum matter is included, Quantum Field Theory and 
String phases appear, and ${\cal R}$-relations between them manifest
as well. These ${\cal R}$-transformations are explicitly shown in two 
relevant examples: Black Hole and de Sitter space times.

\end{abstract}

\footnotetext[2]{Observatoire de Paris, Demirm (Laboratoire Associ\'e au CNRS UA 
336, Observatoire de Paris et Ecole Normale Sup\'erieure), 61 Avenue de l'Observatoire, 75014 Paris,
France.}
\end{center}
\normalsize
\newpage

\newpage

The combined study of Quantum Field Theory (QFT) and String Theory (ST) 
in curved space times - and in the framework of the string analogue model -
allows to go further in the understanding of quantum gravity effects Ref. [1,2,3,4]. Here, QFT and ST are both defined in a non compact D-dimensional background
(although some of the dimensions could eventually be compactified). \\

\medskip
The string ``analogue model'' (or thermodynamical approach) is a suitable 
framework for this purpose.
Strings are considered as a collection of quantum fields $\phi_n$ which do not
interact among themselves but are coupled to the curved 
background, and whose masses are given by the degenerated string spectrum in 
this background. The 
higher mass spectrum is described by the density of string mass levels $\rho (m)$ in the space time
considered Ref. [3,4]. \\

In Black Hole backgrounds (BH), $\rho (m)$ is the same as in flat space time
Ref. [5] and the string critical temperature is the usual (Hagedorn) temperature
in flat space time. In de Sitter (dS) as well as in Anti de Sitter (AdS)
backgrounds (including the corresponding conformal invariant WZWN SL(2,R) models) $\rho (m)$ are different from flat space time Ref. [3,6,7,8,9]. There is a
critical (maximal) string temperature in de Sitter background, while in AdS
there is no finite maximal string temperature at all Ref. [6,7,9] ( the
"maximal" temperature is formally infinite, as the partition function for a gas
of strings in AdS is defined for any external temperature Ref. [6]).

\medskip
We define a ${\cal R}$ ``dual'' transformation over a length $L$, as following \\

\begin{equation} 
\tilde{L} = {\cal R} L = {\cal L}_{\cal R} \: L^{-1}                    %(1)
\end{equation}

\bigskip
\noindent
where ${\cal L}_{\cal R}$ has dimensions of $({\rm length})^2$.

\bigskip
For physical theories, ${\cal R}$ maps classical length scales $L_{cl}$ into quantum string length 
scales $L_q$, and conversely \\

$$\begin{array}{lcccr}
& \tilde{L}_{cl} & \equiv & {\cal R} L_{cl} = {\cal L}_{\cal R} \: L^{-1}_{cl} = \: L_q  & \qquad 
\qquad \qquad \qquad \qquad \quad  (2) \\    
{\rm and} \qquad \qquad \qquad \qquad \qquad \qquad \qquad & & & & \\
& \tilde{L}_q & \equiv & {\cal R} L_q = {\cal L}_{\cal R} \: L^{-1}_q = \: L_{cl}  & \qquad \qquad 
\qquad \qquad \qquad \quad (3) 
\end{array}$$

\stepcounter{equation}
\stepcounter{equation}

\bigskip
${\cal L}_{\cal R}$ depends on the
dimensional parameters of the theory and on the string constant $\alpha^{\prime}$ (and on $\hbar$, $c$ as well). 
The ${\cal R}$ transform, as we will show here, is not an assumed or a priori
imposed symmetry, but it is a transformation revealed from the QFT and ST dynamics
in curved backgrounds. \\

%\setcounter{equation}{2}
%\begin{equation}
%{\cal L}_{\cal R} \equiv  L_{cl} \: L_q                         %(3)
%\end{equation}

\bigskip
\noindent
$L_{cl}$ sets up a length scale for the semiclassical - QFT regime and $L_q$ is the length that 
characterizes the string domain; $L_q$ depends on the dimensional string constant $\alpha^{\prime}$
($\alpha^{\prime} = c^2/2 \pi T$, where $T$ is the string tension) and on the specific background considered. \\

The "vehicle" of the ${\cal R}$-transformation is the dimensional constant
$\alpha^{\prime}$ introduced by ST, which is essential here. The 
${\cal R}$-operation transforms the characteristic lengths of one regime (QFT or
ST) into the characteristic lengths of the other, and thus relate 
different physical regimes in the same curved space time. \\

The ${\cal R}$-transformation belongs to the class of relativistic quantum type of operations $L \longrightarrow L^{-1}$, which appear in ST due to the 
existence of the dimensional string constant $\alpha^{\prime}$. (Let us recall
that the T-duality, widely studied in ST Ref. [12], is e.g. $R_1 R_2 = 
{ \alpha^{\prime} \hbar \over c }$, but where $R_1$ and $R_2$ are 
compactification radii. Such a T-duality links physically equivalent string
theories). As  ST should contain QFT in the limit $\alpha^{\prime} \longrightarrow 0$, our study supports the idea of considering  this type of transformations
as fundamental symmetries for the full String Theory.

\medskip
QFT in curved backgrounds with event horizons posseses an intrinsic (Hawking-Gibbons) temperature $T_H$
Ref. [1,10,11], which can be expressed, in general, as a function $T$ of $L_{cl}$ (and of the constants $\hbar$, $c$ and $k_B$) \\

\begin{equation}
T_H = T (L_{cl})                                                %(4)
\end{equation}

\bigskip
Quantum strings in Minkowski space time have an intrinsic (Hagedorn) temperature. Quantum strings in 
curved backgrounds have also an intrinsic string temperature $T_S$ Ref. [2,3,6], which depends on 
$L_q$ \\

\begin{equation}
T_S = T (L_q)                                                   %(5)
\end{equation}

\bigskip
\noindent
Explicit calculations for de Sitter (dS) and Schwarzschild Black Hole (BH) show
that $T$ is formally the same function for both QFT and QS temperatures 
([Eq. (4)] and [Eq. (5)]), Ref. [3,4]. \\

\medskip
It is worth to point out that space times without event horizons -- such as anti de Sitter (AdS) space
-- have neither $T_H$ nor $T_S$ temperatures, independently of how much quantum matter is present. In fact, in pure AdS space, $T_H$ is zero and $T_S$ is formally 
infinite Ref. [6,9]. \\

\medskip
Applying the ${\cal R}$ operation ([Eq. (2)] to [Eq. (4)]) and [Eq. (5)], we read \\

\begin{equation}
\tilde{T}_H = T_S \: \: \: \: , \tilde{T}_S = T_H                      %(6)
\end{equation}

\bigskip
That is, $T_H$ and $T_S$ are mapped one into the other. From the above equations, we can also write \\

\begin{equation}
\tilde{T}_H \: \tilde{T}_S  = T_S T_H                            %(7)
\end{equation}

\bigskip
\noindent
which is a weaker ${\cal R}$-relation between $T_H$ and $T_S$. \\

\medskip
Let us analyse the mass domains in the corresponding QFT and QS regimes in a curved space time. They
will be limited by the corresponding mass scales $M_H$ and $M_{QS}$. $M_H$ depends on the classical 
length $L_{cl}$ \\

\begin{equation}
M_H = M (L_{cl})                                                 %(8)
\end{equation}

\bigskip
\noindent
and $M_{QS}$ on the quantum length $L_q$, through the same formal relation as we
shall see\\

\begin{equation}
M_{QS} = M (L_q)                                                 %(9)
\end{equation}
 
\bigskip
Therefore, under the ${\cal R}$ operation [Eq. (2)] the mass scales satisfy: \\

\begin{equation}
\tilde{M}_H = M_{QS} \: \: \: \: , \tilde{M}_{QS} = M_H                 %(10)
\end{equation}

\bigskip
On the other hand, if $m_{QFT}$ is the mass of a test particle in the QFT regime and $m_S$ the mass of
a particle state in the quantum string spectrum, the mapping of the QFT mass domain ${\cal D}$ on the
QS domain -- and viceversa -- will read \\

\begin{equation}
{\cal R} \left( {\cal D} ( m_{QFT} , M_H ) \right) = {\cal D} (m_S , M_{QS} )       %(11)
\end{equation}

\bigskip
Summarizing: The QFT regime -- characterized by $L_{cl}$, $T_H$ and $M_H$ -- and the QS regime -- 
characterized by $L_q$ , $T_S$ and $M_{QS}$ -- , both in a curved space time, are mapped one into
another under the ${\cal R}$-transform.  \\

\medskip
We illustrate this with two relevant examples: de Sitter (dS) and Black Hole (BH) space times. \\

\bigskip \bigskip \bigskip

$1$. {\Large \underline {QFT and QS in de Sitter space time}}

\bigskip \bigskip
The classical $L_{cl}$, or horizon radius, is \\

\begin{equation}
L_{cl} = c H^{-1}                                               %(12)
\end{equation}

\bigskip
\noindent
where $H$ is the Hubble constant. The mass scale $M_H$ is such that $L_{cl}$ is its Compton wave 
length \\

\begin{equation}
M_H = \: \frac{\hbar}{c L_{cl}} \: = \: \frac{\hbar H}{c^2} \qquad \qquad \qquad \qquad ,     %(13)
\end{equation}

\bigskip
\noindent
and the QFT Hawking-Gibbons temperature is \\

\begin{equation}
T_H = \: \frac{\hbar}{2 \pi k_B c} \:  \kappa                   %(14)
\end{equation}

\bigskip
\noindent
here $\kappa$ is the surface gravity. For dS space time $\kappa = cH$, and $T_H$ reads \\
 
\begin{equation}
T_H = \: \frac{\hbar H}{2 \pi k_B} \: = \frac{\hbar c}{2 \pi k_B} \: \left( \frac{1}{L_{cl}} \right)
                                                                 %(15)
\end{equation}

\bigskip
On the other hand, canonical as well as semiclassical quantization of quantum strings in dS space time
Ref. [6,7,8] lead to the existence of a maximum mass $m_{\max}$ for the quantum string (oscillating or
stable) particle spectrum. We identify this maximal mass with the string mass scale $M_{QS}$ \\

\begin{equation}
M_{QS} \equiv m_{\max} \simeq c ( \alpha^{\prime} H )^{-1}                    %(16)
\end{equation}

\bigskip
The fact that there is a maximal mass $M_{QS}$ implies the existence of a (minimal) quantum string 
scale $L_q$, which is the corresponding Compton wave length \\

\begin{equation}
L_q = \: \frac{\alpha^{\prime} \hbar H}{c^2} \: = \: \frac{\hbar}{c M_{QS}}  \qquad
\qquad \qquad \qquad ,                                                          %(17)
\end{equation}

\bigskip
\noindent
and of a maximum (or critical) temperature $T_S$ \\

\begin{equation}
T_S = \: \frac{\hbar c}{2 \pi k_B} \: \left( \frac{1}{L_q} \right) \: = \: \frac{c^3}{2 \pi k_B
\alpha^{\prime} H}                                                             %(18)
\end{equation}

\bigskip
Notice that $L_q$, $T_S$ and $M_{QS}$ depend on the string tension and on $H$  while 
${\cal L}_{\cal R}$ [Eq. (3)] does only on $\alpha^{\prime}$:

\begin{equation}
{\cal L}_{\cal R} = \alpha^{\prime} \hbar c^{-1} \equiv L^2_S                  %(19)
\end{equation}

\bigskip
\noindent
$L_S$ is a pure string scale. \\

\medskip
From the above equations it is clear that the ${\cal R}$-transform maps the set 
($L_{cl}$, $T_H$ and $M_H$) into the other set ($L_q$, $T_S$
and $M_{QS}$), satisfying the duality relations [Eqs. (2,6 and 10)]. \\

\medskip
We consider now the mass domains in the two regimes. In the QFT regime the mass $m_{QFT}$ of a quantum
test particle satisfies $m_{QFT} < M_H$. In the QS regime, the mass $m_S$, of a particle state from the
quantum string spectrum satisfies $m_S < M_{QS}$. These conditions can be expressed in terms of 
lengths as $\lambda_{QFT}/L_{cl} > 1$ and $\lambda_S/L_q > 1$, where $\lambda_{QFT}$ and $\lambda_S$
are the Compton wave lengths corresponding to $m_{QFT}$ and $m_S$ respectively. The ${\cal R}$ 
operation then maps the QFT domain into the QS domain [Eq. (11)] \\

\begin{equation}
{\cal R} \: \left( \frac{m_{QFT}}{M_H} \right) \: = \: \frac{m_S}{M_{QS}} \: < 1          %(20)
\end{equation}

\bigskip
On the other hand, string quantization in dS space time requires $L_q$ to be much smaller than the 
Universe radius i.e $L_q/L_{cl} \ll 1$. As a consequence, $M_H$ and $M_{QS}$ obey the inequality
$M_{QS} \gg M_H$.

\bigskip
If back reaction effect of the quantum matter is considered, 
${\cal R}$-relations between the QFT and ST regimes manifest as well. We
studied quantum string back reaction due to the higher excited modes in the 
framework of the string
analogue model. Two branches $R_{\pm}$ (i.e. $H_{\pm}$) of solutions for the 
scalar curvature show up Ref. [3]:

\begin{description}
\item[(a)]
A high curvature solution $R_+$ with a maximal value $R_{\max} = (9 c^4 \pi^2 /4G) (6/(5 
\alpha^{\prime} c \hbar^3))^{1/2}$, entirely sustained by the strings. This branch corresponds to the 
string phase for the background, whose temperature is given by the intrinsic 
string temperature \\

\begin{equation}
T_S \equiv T_+ = \: \frac{c^3}{2 \pi k_B \alpha^{\prime} H_+} \qquad \qquad \qquad \qquad \cdot  %(21)
\end{equation}

\medskip

\item[(b)]

A low curvature solution $R_-$ whose leading term in ${\cal R}/{\cal R}_{\max}$ expansion is the
classical curvature. This branch corresponds to the semi-classical-QFT phase of the background geometry, whose
temperature is given by the Hawking-Gibbons temperature \\

\begin{equation}
T_H \equiv T_- = \: \frac{\hbar H_-}{2 \pi k_B}                                            %(22)
\end{equation}

\end{description}
\bigskip
From [Eq. (21)] and [Eq. (22)], we see that the ${\cal R}$-relations 
manifest as well when back reaction is included.

\bigskip \bigskip \bigskip
 
$2$. {\Large \underline{QFT and QS in Schwarzschild BH space time}}

\bigskip \bigskip
Here $L_{cl}$ is the BH radius $r_H$ ( $G$: gravitational Newton constant):

\begin{equation}
r_H = \: \left( \frac{16 \pi G M_H}{c^2 (D-2) A_{D-2}} \right)^{\frac{1}{D-3}} \qquad \qquad , \: 
\left( A_{D-2} \equiv \: \frac{2 \pi^{\frac{(D-1)}{2}}}{\Gamma \left( \frac{D-1}{2} \right)} \right) 
                                                                                            %(23)
\end{equation}

\bigskip
\noindent
and $M_H$ is the BH mass. \\

\medskip
The QFT Hawking temperature is \\

\begin{equation}
T_H = \: \frac{\hbar c (D-3)}{4 \pi k_B} \: \left( \frac{1}{r_H} \right)               %(24)
\end{equation}

\bigskip
\noindent
($\kappa = (D-3) c^2 / 2 r_H$).

\bigskip
In the QFT regime, a BH does emit thermal radiation at a temperature $T_H$ Ref. [10]. In the QS regime,
the BH has a high massive quantum thermal emission, corresponding to the higher excited quantum string
states. For open strings (in the asymptotically flat BH region), the thermodynamical behaviour of these
states is deduced from the string canonical partition function Ref. [4]. In BH space times, the mass
spectrum of quantum string states coincides with the one in Minkowski space, and critical dimensions 
are the same as well, Ref. [5]. Therefore, the asymptotic string mass density of levels, in BH space 
times, reads $\rho (m) \sim \exp \{ b (\alpha^{\prime} c / \hbar )^{1/2} m \}$ , and quantum strings 
have an intrinsic temperature $T_S$ which is the same as in flat space time. The string canonical 
partition function is defined for Hawking temperatures $T_H$ satisfying the condition Ref. [4] \\

\begin{equation}
T_H  < T_S = \: \frac{\hbar c}{b k_B L_S}                            %(25)
\end{equation}

\bigskip
\noindent
i.e in string theory, $T_H$ has an upper limit given by the intrinsic or critical string temperature 
$T_S$. This limit implies the existence of a minimal BH radius $r_{\min}$, and a minimal BH mass
$M_{\min}$ \\

$$\begin{array}{lclr}
& r_H > r_{\min} = & \: \frac{b (D-3)}{4 \pi} \: L_S  & \qquad \qquad  \qquad   (26)        \\           
& &  & \\
{\rm and} \qquad \qquad \qquad \qquad \qquad \qquad  & & & \\
& & &  \\
& M_H > M_{\min} = & \: \frac{c^2 (D-2) A_{D-2}}{16 \pi G} \: \left( \frac{b (D-3) L_S}{4 \pi} 
\right)^{D-3} & \qquad \qquad \qquad  (27)                                                    
\end{array}$$
\bigskip
\noindent
Here $L_S$ is given by [Eq. (19)], and ${\cal L}_R$ [Eqs.(2,3)] is given by

$${\cal L}_R={b M \alpha^{\prime {3 \over 2}} \over 2 M } \sqrt{{\hbar \over c}}, \qquad D=4\,. $$

\bigskip
Therefore, the QS scales are $L_q = r_{\min}$ , $M_{QS} = M_{\min}$ , and the string temperature $T_S$.
They depend on the type of strings and on the dimension through the parameter $b$. In terms of 
$r_{\min}$, $T_S$ and $M_{\min}$ read \\

$$\begin{array}{lcccr}
& T_S  & = & \: \frac{\hbar c (D-3)}{4 \pi k_B} \: \left( \frac{1}{r_{\min}} \right) & \qquad 
\qquad \qquad  \qquad \qquad   \qquad \qquad (28) \\
\qquad \qquad \qquad \qquad \qquad \qquad & & & & \\
& M_{\min} & = & \frac{c^2 (D-2) A_{D-2}}{16 \pi G} \: r^{D-3}_{\min}  & \qquad \qquad \qquad 
\qquad \qquad \qquad  (29)                    
\end{array}$$

\bigskip
\noindent
From [Eqs. (23,24)] and [Eqs. (28,29)], we see that ($r_H , M_H , T_H$) and ($r_{\min} , M_{\min} ,
T_S$) are ${\cal R}$-transformed of each other. This is valid in all dimensions. \\

\bigskip
We discuss now the mass domain relation [Eq. (11)] in the BH case. The QFT domain reads $\lambda_{QFT}
/r_H < 1$, and the QS domain $\lambda_S /r_{\min} < 1$. In terms of the masses involved ( $D = 4$, for
simplicity), these conditions yield to $M^2_{PL} (M_H m_{QFT} )^{-1} < 1$ and $M^2_{PL} (M_{\min} m_S 
)^{-1} < 1$ , ($M_{PL} = (\hbar c/G )^{1/2}$). Both mass conditions can be expressed as $\beta_H
m_{QFT} c^2 > 1$ and $\beta_S m_S c^2 > 1$, where $\beta = (k_B T)^{-1}$ being $T$ the BH temperature 
in each regime. This is precisely the condition for defining the asymptotic expression for the string
canonical partition function Ref.[4]. \\

\medskip
The physical meaning behind these ${\cal R}$-relations is the following. At the first stages of BH 
evaporation, emission is in the lighter particle masses at the Hawking temperature $T_H$, as described
by the semiclassical - QFT regime. As evaporation proceeds, temperature increases and high massive 
emission
corresponds to the higher excited string modes. At the later stages, for $T_H \to T_S$ (i.e $r_H \to
r_{\min}$ , $M_H \to M_{\min}$), the BH enters its QS regime. The ${\cal R}$ 
transformation allows to link
the early ($T_H \ll T_S$, i.e $r_H \gg r_{\min}$, $M_H \gg M_{\min}$) and late ($T_H \to T_S$ , i.e 
$r_H \to r_{\min}$ , $M_H \to M_{\min}$) stages of BH evaporation. \\

\medskip
These ${\cal R}$-relations manifest as well if the back reaction effect of higher massive string modes is included
Ref. [4]. The string back reaction solution ($r_+$, $M_+$, $T_+$) shows that the BH radius $r_+$ and
mass $M_+$ decrease and the BH temperature $T_+$ increases. Here $r_+$ is bounded from below (by
$r_{\min}$) and $T_+$ does not blow up ($T_S$ is the maximal value). The string back reaction effect is
finite and consistently describes both the QFT ($T_H \ll T_S$) and the QS ($T_H \to T_S$) regimes. It 
has the bounds: $(r,T,M)_{\min} < (r,T,M)_+ < (r,T,M)_H$. The two bounds are linked by the ${\cal R}$
transform. \\

\medskip
Summarizing, from the above BH and dS studies, the Hawking-Gibbons temperature of the QFT regime
becomes the intrinsic string temperature of the string regime. \\
Also, in the AdS background, the ${\cal R}$-transformation manifests as well between $T_H$ and $T_S$.

\medskip
These BH and dS examples suggest that our ${\cal R}$ transform could be promoted to a dynamical
operation: evolution from a semiclassical - QFT phase to a quantum string phase (as in BH 
evaporation) or conversely (cosmological evolution).

\vskip 2cm

{\Large \underline{Acknowledgements}}

\bigskip \bigskip

M.R.M acknowledges C.A.I.C.Y.T under project AEN97 -- 1693 for partial financial support and the 
Observatoire de Paris -- DEMIRM for the kind hospitality during this work. \\

Partial financial support from NATO Collaborative GRANT CRG974178 is also acknowledged (N.S).

\newpage

{\Large \underline{References}}

\bigskip \bigskip

1. N.D. Birrell and P.C.W. Davies, {\it Quantum Fields in Curved Space}, ( Cambridge University Press,
England, 1982).
\medskip

2. {\it String Theory in Curved Space Times}, edited by N. S\'anchez (World Scientific. Pub, Singapore,
1998).
\medskip

3. M. Ramon Medrano and N. S\'anchez, Phys. Rev.  D60, 125014 (1999).
\medskip

4. M. Ramon Medrano and N. S\'anchez, Phys. Rev. D61, 084030 (2000).
\medskip

5. H.J. de Vega and N. S\'anchez, Nucl. Phys. B309, 522 (1988); B309, 577 (1988)
.

6. A.L. Larsen and N. S\'anchez, Phys. Rev. D52, 1051 (1995).
\medskip

7. H.J. de Vega, A.L. Larsen and N. S\'anchez, Phys. Rev. D15, 6917 (1995).
\medskip

8. H.J. de Vega and N. S\'anchez, Phys. Lett. B197, 320 (1987).
\medskip

9. H.J. de Vega, L. Larsen, N. S\'anchez, Phys Rev. D58, 026001(1998).
\medskip

10. S. W. Hawking, Comm. Math. Phys. 43, 199 (1975).
\medskip

11. G.W. Gibbons and S. W. Hawking, Phys. Rev. D15, 2738 (1977).
\medskip

12. J. Polchinski, {\it String Theory} (Vol I and II. Cambridge University Press, 1998), and references therein.

\end{document}